\documentclass[journal=jacsat,manuscript=article]{achemso}
\usepackage[version=3]{mhchem} 
\usepackage{amsmath}
\usepackage[T1]{fontenc} 
\usepackage{xcolor}

\def\excLambda{980} 
\def\smallPSsize{40} 
\def\largePSsize{200} 
\def\PTBsize{190} 
\def\maxVel{2.4} 
\def\lowPower{0.785} 
\def\smallPSheight{80} 
\def\relPSheight{10} 
\def\longResAir{850} 
\def\transvResAir{550} 

\author{Marciano Palma do Carmo}
\email{marciano.palma_do_carmo@kcl.ac.uk}
\affiliation{Physics Department, King's College London, London, UK}
\author{David Mack}
\affiliation{Department of Physics, Imperial College London, London, UK}
\author{Diane J. Roth}
\author{Miao Zhao}
\author{Ancin M. Devis}
\author{Francisco J. Rodr\'iguez-Fortu\~no}
\affiliation{Physics Department, King's College London, London, UK}
\author{Stefan A. Maier}
\affiliation{School of Physics and Astronomy, Monash Unversity, Clayton Victoria, Australia}
\alsoaffiliation[Second University]
{Department of Physics, Imperial College London, London, UK}
\author{Paloma A. Huidobro}
\affiliation{Departamento de F\'isica Te\'orica  de la Materia Condensada, Universidad Aut\'onoma de Madrid, Spain}
\alsoaffiliation[Second University]
{Condensed Matter Physics Center (IFIMAC), Universidad Aut\'noma de Madrid, Madrid, Spain}
\altaffiliation{Instituto de Telecomunica\c{c}\~oes, Instituto Superior Tecnico - University of Lisbon, Lisboa, Portugal}
\author{Aliaksandra Rakovich}
\affiliation{Physics Department, King's College London, London, UK}
\email{aliaksandra.rakovic@kcl.ac.uk}

\title[Plasmonic brownian ratchets for directed transport of analytes]
{Plasmonic brownian ratchets for directed transport of analytes}

\keywords{Brownian ratchets, analyte motion, plasmonics, optical trapping}

\begin{document}

\begin{tocentry}

\centering
\includegraphics[width=8.2cm]{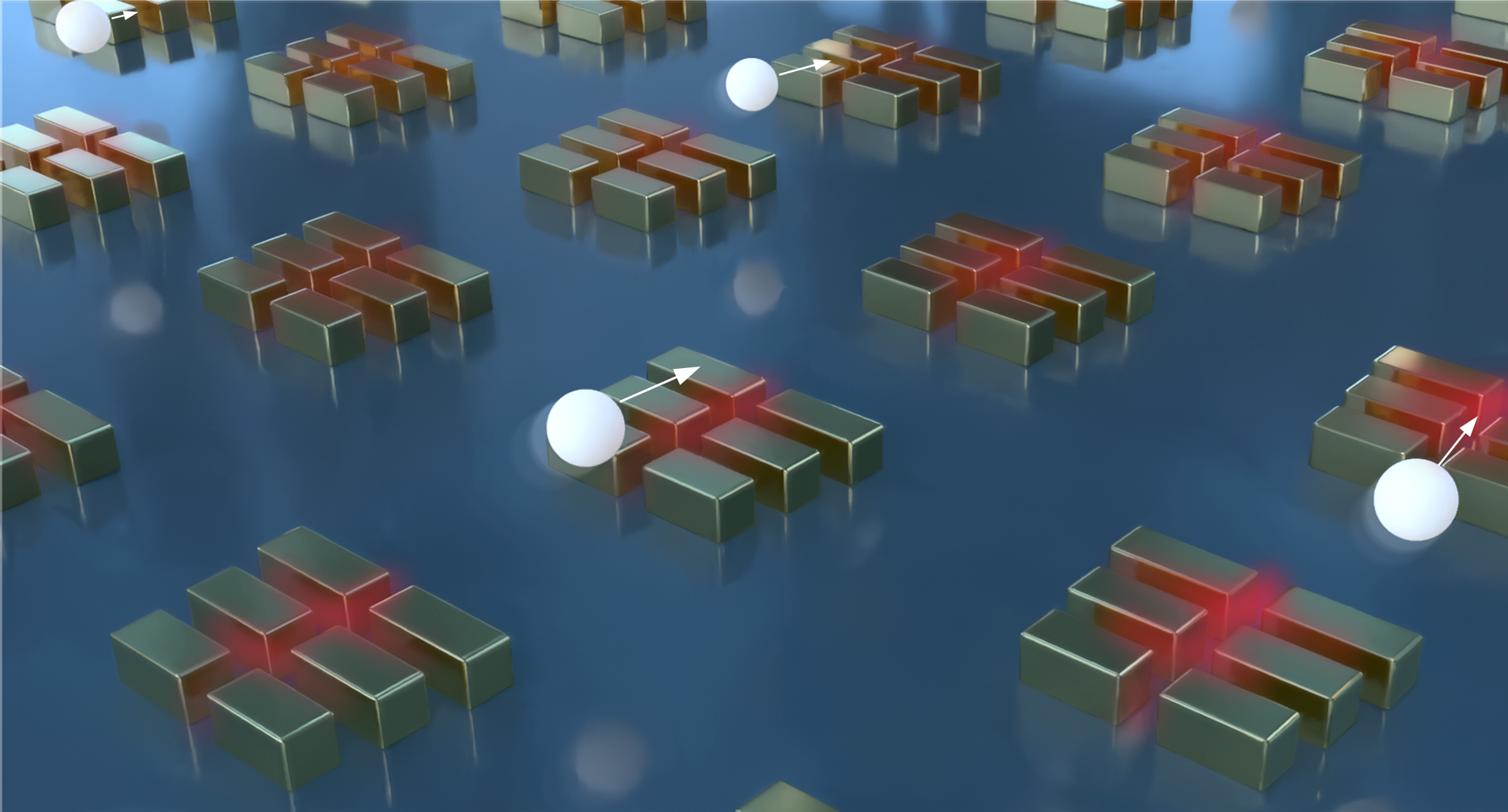}



\end{tocentry}

\begin{abstract}

Plasmonic nanostructures provide strong optical near-fields for trapping and manipulating nano-sized particles, but converting these interactions into robust directional transport has remained challenging. 
Here we demonstrate a plasmonic Brownian ratchet that rectifies colloidal diffusion using an asymmetric gold nanoarray under continuous-wave illumination. 
Finite-element simulations reveal anisotropic near-field distributions that bias optical forces, and experiments confirm directed motion for 40 - 200 nm nanoparticles of various compositions (dielectric, semi-conduction and metallic). 
We show that, under periodic light modulation, nanoparticles undergo unidirectional lateral transport with velocities up to \(\maxVel~\mathrm{\mu m/s}\) at incident intensities below \(1~ \mathrm{kW/cm^{2}}\). 
These results establish plasmonic ratcheting as an efficient route to bias transport of nano-sized analytes, achieving markedly higher speeds and lower operating powers than previous optical ratchets, and opening opportunities for integration into nanophotonic and lab-on-chip systems.

\end{abstract}


Controlled long-range transport of micro- and nanoscale objects underpins many lab-on-a-chip and microfluidic technologies, enabling capture, concentration, manipulation, and detection of analytes. 
Applications range from sorting microorganisms such as bacteria, algae, and blood cells,\cite{FavreBulle2019optical} to stretching DNA and RNA,\cite{svoboda1994biological} to probing molecular motor forces and chromosome sorting,\cite{neuman2004optical,grier2003revolution} cancer cell classification,\cite{paiva2020ilof} and disease diagnostics.\cite{polimeno2018optical} 
Existing approaches rely on microfluidic pumps,\cite{Binsley2020microfluidic,Liu2017high} optical tweezers,\cite{ashkin1987optical} capillary forces,\cite{sanavio2015on,magnasco1998feynman} or Brownian diffusion,\cite{sanavio2015on} but each carries limitations: pumps require high flow rates and power, optical forces are constrained by diffraction and demand high intensities that damage biological analytes, and passive capillary or diffusive transport suffers from low analyte residence times and requires high concentrations.

Nonetheless, the transport of analytes by Brownian motion is extremely appealing because it can occur over long distances and at negligible power cost, without requiring complex fluidic systems or external flows. 
The main drawback of Brownian motion -- its lack of directionality -- can be overcome by rectifying analyte motion through the application of an asymmetric, temporally modulated trapping potential (see Supplementary Information (SI) section ``Principles of Brownian Ratchets'' and Figure S1 for further background). 
This constitutes the principle of a Brownian ratchet which has been widely studied in physics and biology,\cite{astumian1997thermodynamics,REIMANN2002,magnasco1998feynman, rousselet1994directional,bader1999dna,faucheux1995optical,vanOudenaarden1999brownian,lee2005observation,wu2016near} and realized experimentally using mechanical, \cite{magnasco1998feynman,semeraro2023diffusion,spiechowicz2023diffusion} electrical \cite{rousselet1994directional,bader1999dna}, thermal, \cite{faucheux1995optical,astumian1997thermodynamics,vanOudenaarden1999brownian,lee2005observation} and photonic \cite{wu2016near} systems. 
Among these realizations, photonic Brownian ratchets are particularly attractive because optical fields provide tunable, non-contact forces that can be readily integrated into lab-on-chip platforms.
Photonic crystal implementation has demonstrated for directed transport of nanoparticles under asymmetric optical potentials,\cite{wu2016near} but the transport speeds achieved were relatively low ($\leq 1~\mathrm{\mu m~ s^{-1}}$) and required high optical powers. 
However, plasmonics has been proposed as a remedy to these limitations,\cite{huidobro2013plasmonic} but no experimental demonstrations have yet been reported -- a gap that the present work aims to address.

Plasmonic nanostructures support localized surface plasmon resonances (LSPRs), which give rise to strong optical field confinement in their near-field.\cite{liao2023highly,giannini2011plasmonic} 
This concentration of fields has the potential to reduce the powers required to operate a Brownian ratchet compared to photonic counterparts. 
In addition, the geometry and dimensions of plasmonic structures can be tailored to match different analyte types, offering scalability and versatility of the plasmonic ratchet systems.
Building on these advantages, we design, fabricate, and test an asymmetric plasmonic nanoarray that generates the required asymmetric potential landscape, and experimentally demonstrate directed Brownian ratcheting of various types of nanoparticles under chopped optical excitation.

Our plasmonic ratchet design follows the theoretical proposal of Huidobro \textit{et al.},\cite{huidobro2013plasmonic} who showed that asymmetric metallic nanostructures can generate periodic optical potentials with broken inversion symmetry suitable for Brownian rectification -- the asymmetric geometry ensures that, under optical excitation, the LSPRs produce a near-field intensity distribution that lacks mirror symmetry along the transport axis, giving rise to the biased optical forces required for ratcheting of analyte particles.
We adopted the theoretical design proposed in this work, but adapted the unit cell dimensions and periodicity for operation at our experimental wavelength (980 nm) and for compatibility with the sub-200 nm nanoparticles studied here. 

\begin{figure}[ht]
\centering
\includegraphics[width=1.0\textwidth]{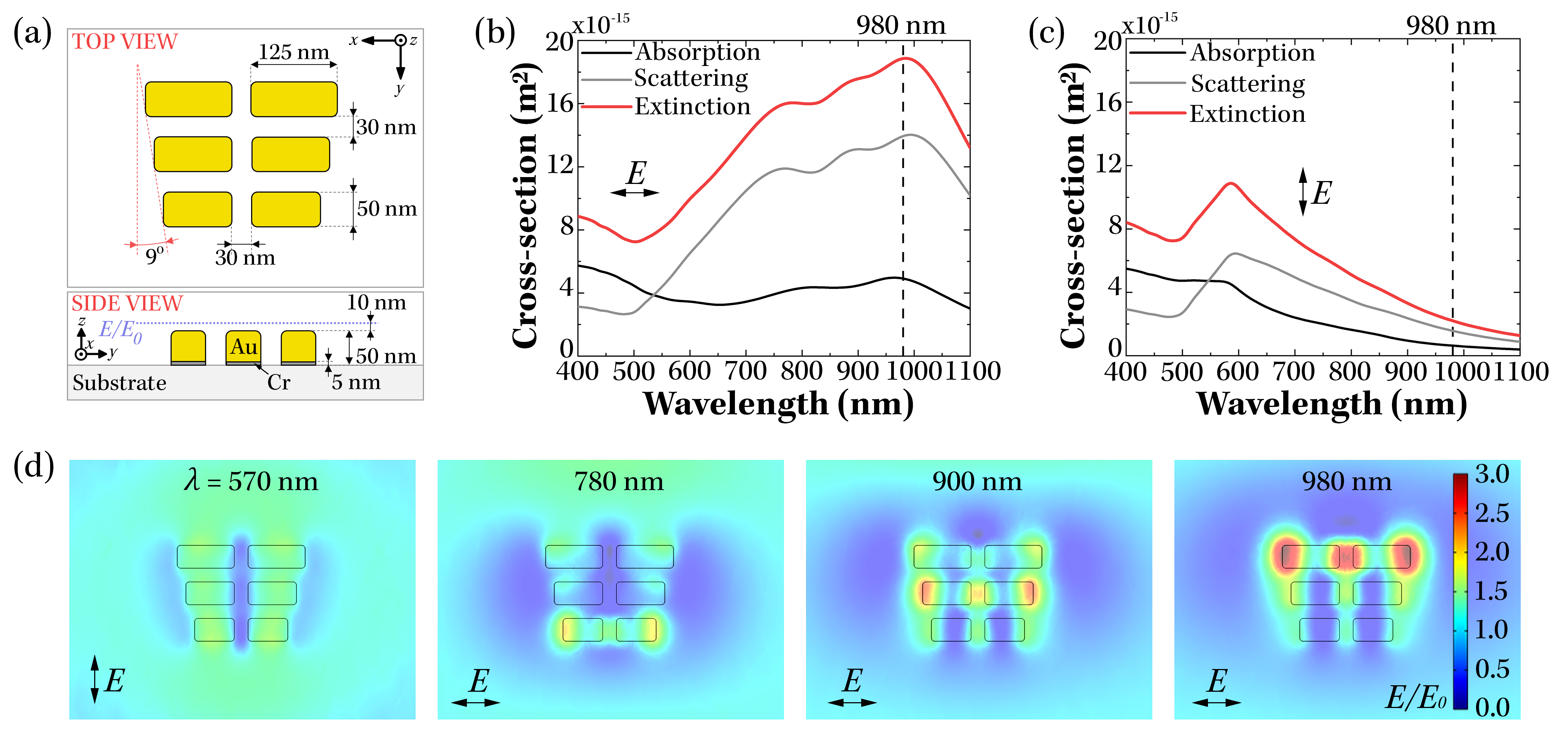}
\caption{Numerical calculations of the optical response of plasmonic ratchets. \textbf{(a)} 
    Optimised structure of the plasmonic ratchet, shown in top view (top box) and in side view (bottom box). The blue dashed line denotes the plane for which the electric field enhancement maps in (d) were calculated. The system of axes shown in panel (a) is maintained throughout the text. \textbf{(b)} and \textbf{(c)} show the numerically simulated absorption (black line), scattering (grey line) and extinction (red line) cross-sections for the ratchet in water, when excited with plane waves of two orthogonal polarisations. The polarisation of electric field is indicated using black arrows in each panel. The black dashed lines in (b) and (c) respresent the wavelength of excitation used in ratchetting experiments (\(\lambda_{exc}=\excLambda~\mathrm{nm}\)). \textbf{(d)}~The calculated electric field enhancements at a plane located \(z=\relPSheight~\mathrm{nm}\) above the top of plasmonic ratchets, for different wavelengths of excitations and different polarisations of incident electric field, as indicated in each plot.}
\label{fig:antena-design}
\end{figure}

Finite-element simulations (COMSOL Multiphysics) were used to optimize the unit cell geometry (see SI section ``Numerical Simulations'' for general methodology).
The optimization was guided by three criteria: (i) the presence of the main plasmonic resonance at the experimental wavelength of 980 nm, (ii) an optical potential depth $\Delta U$ exceeding the thermal energy $k_B T$, and (iii) sufficient potential asymmetry to bias particle diffusion.
In this instance, the optical forces were estimated within the dipolar approximation, which provides a rapid yet reliable description for sub-wavelength particles in the near-field (SI subsection ``Optimization of ratchet asymmetry via dipole approximation'').
To meet the optimization criteria, we first tuned a single gold (Au) dimer by sweeping rod length ($L$), width ($w$), height ($h$), and gap ($g$) dimensions.
The optimized dimer, consisting of two rectangular rods ($L=125$ nm, $w=50$ nm, $h=30$ nm; 5 nm Cr adhesion layer on glass + 45 nm Au) separated by a $\approx 30$ nm gap, exhibited a longitudinal resonance near the target wavelength of 980 nm, in water (see Figure~\ref{fig:antena-design}).

Building on the Huidobro design,\cite{huidobro2013plasmonic} the full unit cell was then constructed from three dimer-antennas with rod lengths ($L$) decreasing along the transport axis, introducing a structural asymmetry quantified by the tilt angle $\theta$ defined by the edges of the rods in the three rows, as can be seen in Figure 2(a).
We performed a parameter sweep of $\theta$ over the range $0-15^\circ$ (Figure S2) and found that $\theta\approx 9^\circ$ maximized both near-field asymmetry and its spatial extent (Figure 2(d)).
Spectral calculations confirmed three longitudinal resonances at $\approx780$, 900, and 980 nm under longitudinal polarization, and a transverse mode at 570 nm (Figure 2(b)-(d)).

\begin{figure}[ht]
\centering
\includegraphics[width=1.0\textwidth]{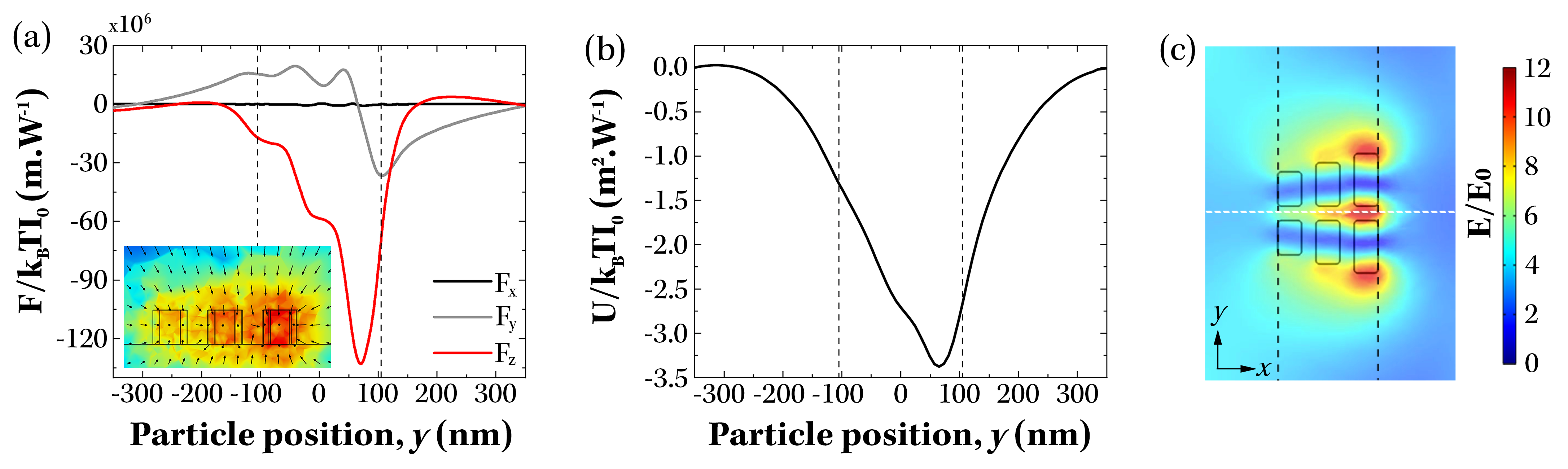}
\caption{Numerical simulations of trapping ability of plasmonic ratchets. The calculated \textbf{(a)} optical forces and \textbf{(b)} the trapping potentials for a \(\smallPSsize~\mathrm{nm}\) PS nanoparticle located at a distance of \(z=\smallPSheight~\mathrm{nm}\) above the array of brownian ratchets, as a function of position \(y\) along the ratchet. The inset of panel (a) shows the stress tensor calculated for \(x=0\) plane. \textbf{(c)} Shows the electric field enhancement at \(z=\smallPSheight~\mathrm{nm}\) plane above the ratchet and the definition of \(x\) and \(y\) directions relative to the orientation of the ratchet. Results shown in panels (a) and (b) were calculated along the dashed white line in this map (corresponds to \(x=0~\mathrm{nm}\) and \(z=\smallPSheight~\mathrm{nm}\) direction). The black dashed lines in all panels represent the edges of the plasmonic ratchets along the \(y\)-direction.}
\label{fig:forcespots_last.png}
\end{figure}

To extend beyond single-unit simulations, the one-dimensional and two-dimensional arrays of unit cells were modeled by first applying periodic boundary conditions along the intended transport direction ($y$) and then also in the transverse direction in the plane of the array ($x$).
Optical forces were calculated using the full Maxwell stress tensor (MST) formalism, as we expected that the intended analyte nanoparticles (sub-200 nm) would not be negligible compared to the array periodicity and their polarizability would significantly perturb the near-fields generated by the plasmonic arrays (see SI section ``Maxwell Stress Tensor calculations of optical forces and potentials'' for further details).
The MST approach therefore provided a rigorous description of the field–particle interaction that could not be captured by the dipolar approximation.
For these calculations, a 40 nm polystyrene (PS) test sphere was positioned 10 nm above the top surface of the ratchet (\textit{i.e.} along $z$ direction).
The optimized array periods were determined to be 700 and 800 nm along the $y$ and $x$ directions respectively (see SI for details of these calculations). 
The resulting force distributions (Figure 3(a)) showed strong attraction in the $z$ direction and a net bias along the $y$ direction, with negligible component in $x$.
Integration of forces along the transport direction yielded an optical potential that was deeper than the thermal energy of the system ($|\Delta U| \simeq 3.4~k_b T$), indicating that the designed structures were capable of efficiently trapping the PS spheres. 
Furthermore, the potential was clearly asymmetric, with the minimum off-set from the geometric centre of the ratchet unit cell by $\Delta y\simeq 65~\mathrm{nm}$, providing the bias required to rectify the Brownian motion of PS spheres;  this asymmetry means that particles must diffuse shorter distances to reach the adjacent potential well in the forward direction than in the backward direction.
Using the Stokes–Einstein diffusion coefficient for 40 nm PS spheres in water, these distances correspond to characteristic diffusion times of $\tau_{F} \approx 4.6$ ms and $\tau_{B} \approx 6.7$ ms at 298 K (SI section ``Principles of Brownian Ratchets'', Equations S1-S2).
Brownian ratcheting is therefore expected only when the off-time of the potential, $\tau_{off}$, lies between these two limits: if $\tau_{off}$ is too short, particles cannot diffuse out of the potential well, while if it is too long, forward and backward diffusion become equally probable and rectification is lost. 
This condition defines the frequency window of optical potential modulation over which directed transport should occur.

\begin{figure}[ht]
\centering
\includegraphics[width=1.0\textwidth]{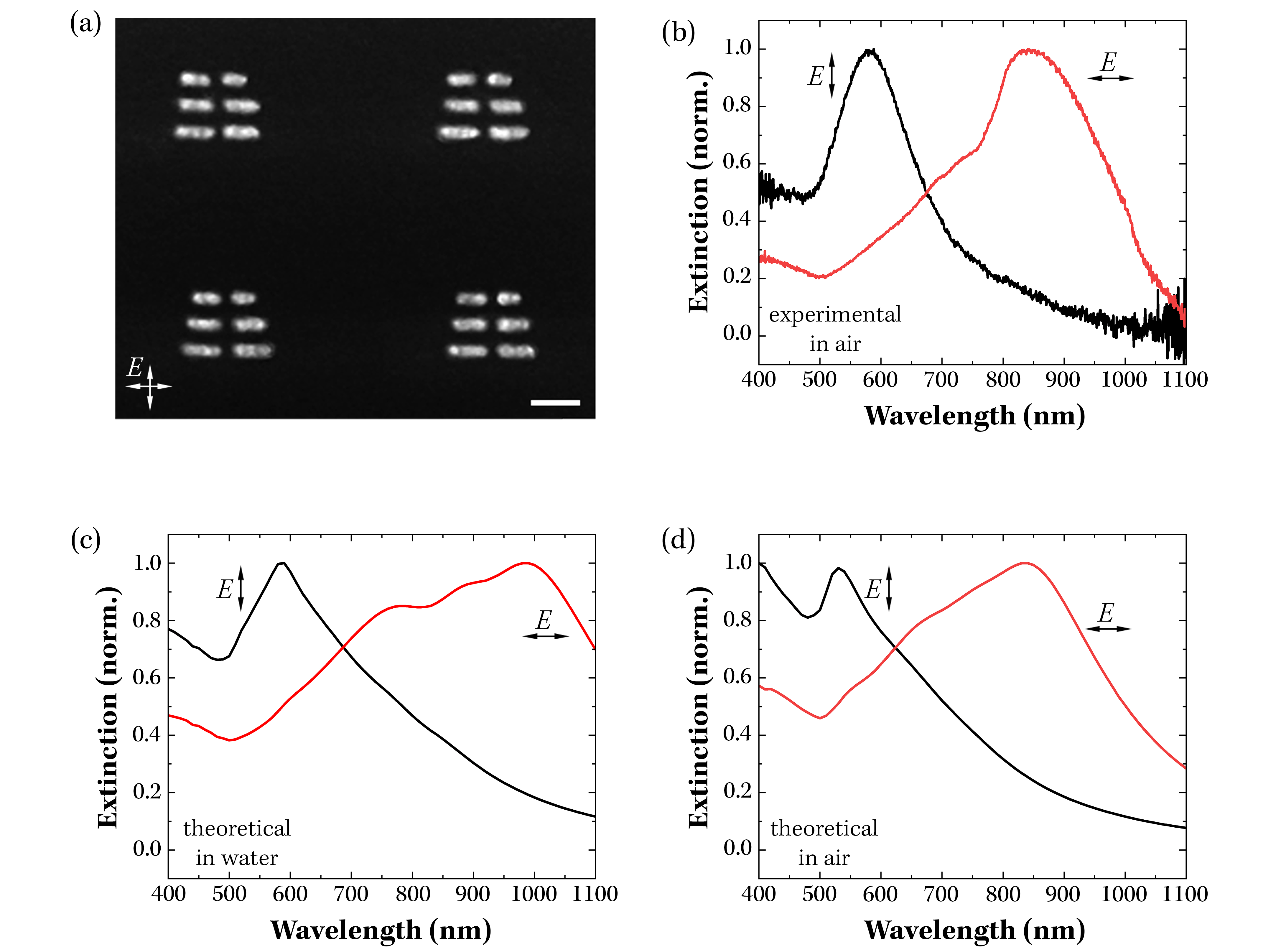}
\caption{Fabrication and optical characterization of plasmonic ratchets. \textbf{(a)} Scanning electron microscope (SEM) image of the fabricated ratchet array. The scale bar in the image corresponds to \(150~\mathrm{nm}\). \textbf{(b)} Experimental normalized extinction cross-sections measured for the plasmonic ratchet array in air. \textbf{(c)} and \textbf{(d)} show the numerically simulated and normalized extinction cross-sections of the ratchet array in water and in air, respectively.}
\label{fig:SEM-and-CSs.png}
\end{figure}

Plasmonic ratchets were fabricated using ebeam lithography (see SI section ``Fabrication protocols'' for details). 
SEM microscopy of representative nanostructures (Figure~\ref{fig:SEM-and-CSs.png}(a)) confirmed that the geometry of the fabricated ratchets was close to the target optimal values ($L = 121 \pm 5$~nm, $g = 32.7 \pm 3.9$~nm, $\theta = 12.3^\circ \pm 1.6^\circ$), although the rod widths ($41.3 \pm 2.5$ nm) were slightly smaller than the design and resulted in correspondingly larger inter-row gaps ($39.8 \pm 2.5$ nm).  
The structures nonetheless exhibited clear asymmetry and were therefore expected to generate the required asymmetric potential for PS sphere diffusion.

Optical characterization of the arrays by extinction spectroscopy at normal incidence in air confirmed that the fabricated ratchets could be resonantly driven at $\excLambda~\mathrm{nm}$. 
The measured spectra (Figure~\ref{fig:SEM-and-CSs.png}(b)) showed excellent agreement with numerical simulations (Figure~\ref{fig:SEM-and-CSs.png}(d)), with both displaying peaks at $\sim\longResAir$~nm and $\sim\transvResAir$~nm corresponding to the longitudinal and transverse resonances of the longest gap-antennas in air. 
Simulations of the arrays immersed in water predicted a red-shift of these resonances (Figure~\ref{fig:SEM-and-CSs.png}(c)) towards the target excitation wavelength of $\excLambda~\mathrm{nm}$.

\begin{figure}[ht]
\centering
\includegraphics[width=1.0\textwidth]{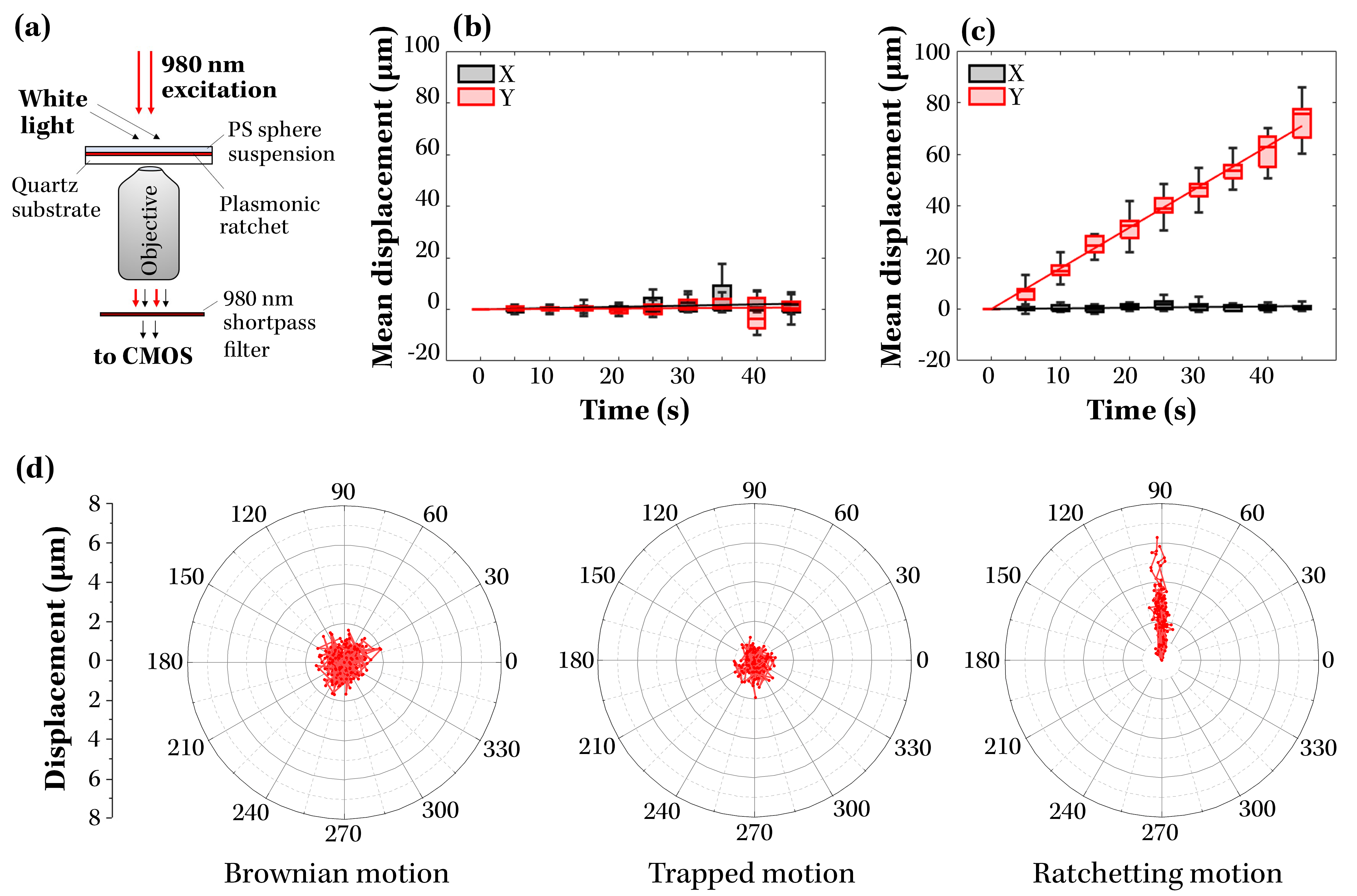}
\caption{Motion of \(\smallPSsize~\mathrm{nm}\) PS spheres under different illumination conditions. Panel \textbf{(a)} shows the set-up used for recording the motion of PS spheres above the plasmonic ratchets.  Panels \textbf{(b)} and \textbf{(c)} compare the average \(x\) and \(y\) displacements of PS particles undergoing Brownian motion and ratchetting, respectively, as a function of motion time. Panel (c) shows, left-to-right, traces of 10 to 20 PS particles undergoing Brownian, trapping or ratchetting motion over a time period of \(270~\mathrm{ms}\). Traces are shown in polar coordinates, with $90^\circ$ direction corresponding to the ratcheting axis ($y$ direction in the text) and with distances measured relative to the position of each particle when first detected.}
\label{fig:ratchetting-experiments}
\end{figure}

To experimentally test the performance of the plasmonic ratchets, the motion of nanoparticles dispersed in water above the arrays was monitored.
The excitation was provided by a $\excLambda~\mathrm{nm}$ CW laser incident normally onto the sample, while the particle motion was imaged in dark-field configuration using side white-light illumination and an objective below the substrate (see schematic in Figure~\ref{fig:ratchetting-experiments}(a)).
Prior to each experiment the $\excLambda~\mathrm{nm}$ laser was blocked to record Brownian motion of the particles to confirm the absence of drifts due to sample tilt or thermal gradients.
The laser power was then increased until stable trapping was observed.
To initiate ratcheting, the excitation was periodically modulated using a 10-slot chopper with 50\% duty cycle, giving an off-time of $\tau_{off}=1/(20f)$, where $f$ is the chopping frequency.
Particle motion at each stage was recorded with a sCMOS camera, and the resulting videos were analyzed to identify and track individual spheres; trajectories were then statistically averaged to yield mean displacement \textit{versus} time data for large ensembles (see Supplementary Information section ``Particle tracking'' for further details).

A subset of trajectories recorded for \(\smallPSsize~\mathrm{nm}\) PS spheres in water is shown in Figure~\ref{fig:ratchetting-experiments}(d) for Brownian, trapping, and ratcheting motion.
For Brownian motion and trapped regimes, the trajectories showed no preferential direction, consistent with the absence of net statistical displacement in both cases; the magnitude of displacements was significantly smaller for trapped particles than for freely diffusing ones, as expected.
In contrast, ratcheted particles exhibited a strong directional bias, displacing predominantly along the ratchet axis (90$^\circ$ in Figure~\ref{fig:ratchetting-experiments}(d), corresponding to the $x$ axis in Figures~\ref{fig:antena-design} and \ref{fig:forcespots_last.png}).
Statistical averages of the particle displacements, presented in Figure~\ref{fig:ratchetting-experiments}(b,c), highlighted the absence of net motion in Brownian diffusion and their pronounced, directed displacement under ratcheting conditions.

\begin{figure}[ht]
\centering
\includegraphics[width=1.0\textwidth]{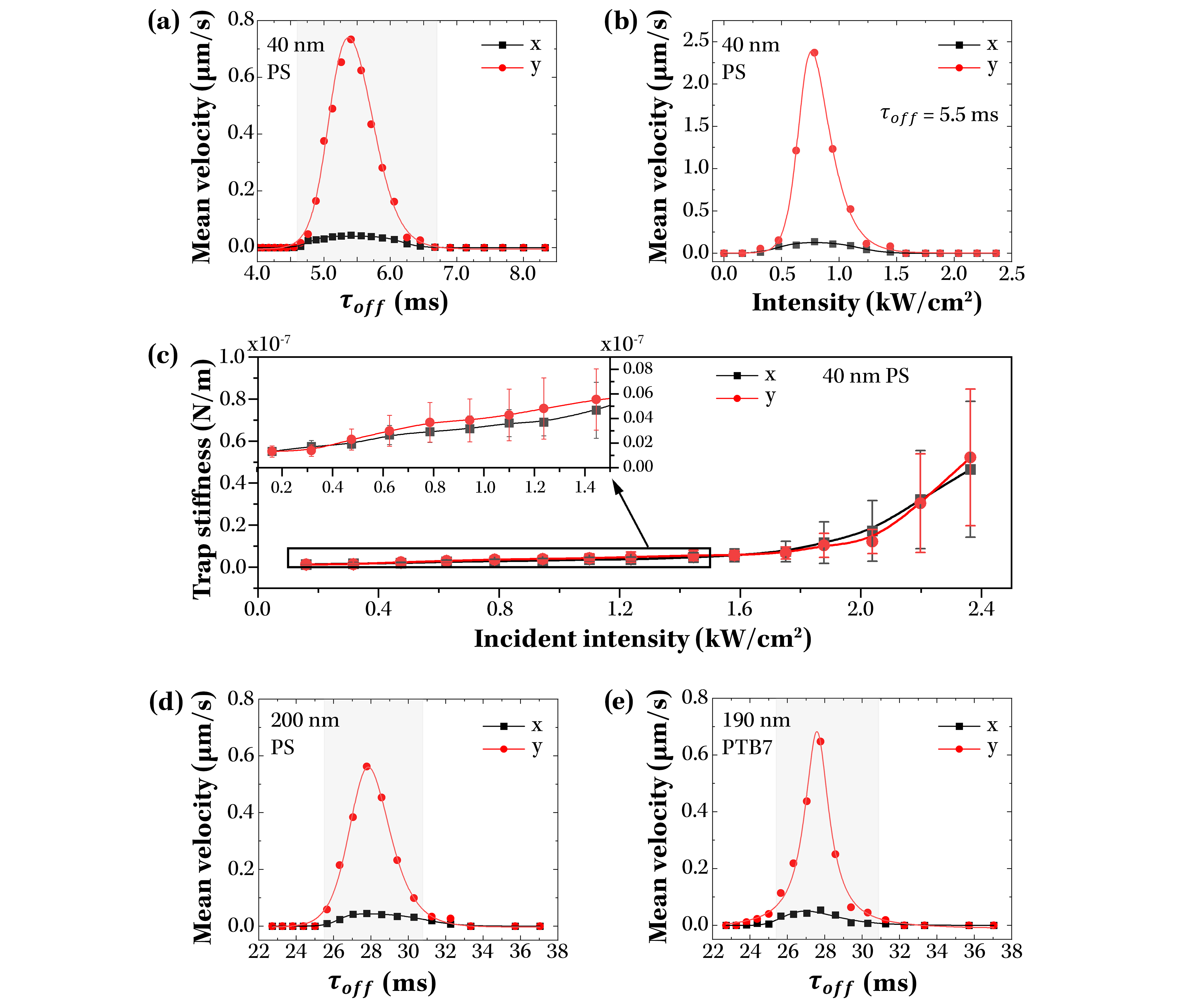}
\caption{The mean velocities achieved during ratcheting of \(\smallPSsize~\mathrm{nm}\) PS spheres as a function of \textbf{(a)} the time the periodic potential remained off (\(t_{off}\)) and \textbf{(b)} the incident excitation intensity, for \(t_{off} = 5.5~\mathrm{ms}\). In panel (a), the incident excitation intensity was $0.47~\mathrm{kW~cm^{-2}}$ and the shaded region indicates the theoretically-predicted range of $\tau_{off}$ values where ratchetting of \(\smallPSsize~\mathrm{nm}\) PS spheres could be achieved. Ratchetting characteristics. (c) Trap stiffness of plasmonic structures under CW illumination. Inset shows data obtained at low incident intensities. (d) and (e) show the mean velocities achieved during ratcheting of \(\largePSsize~\mathrm{nm}\) PS and PTB7 spheres as a function of time the periodic potential remained off (\(\tau_{off}\)). In panel (a), the incident excitation intensity was $0.47~\mathrm{kW~cm^{-2}}$, whereas for panels (d) and (e), the excitation intensities were approximately \(2.5~\mathrm{kW~cm^{-2}}\). The shaded regions in panels (a), (d) and (e) indicate the theoretically-predicted range of $\tau_{off}$ values for successful ratchetting of particle of specified sizes and compositions.}
\label{fig:ratchetting-results}
\end{figure}

Figure~\ref{fig:ratchetting-results}(a) shows the net particle motion resolved along the transverse ($x$) and ratchet ($y$) directions under $\excLambda~\mathrm{nm}$ excitation chopped at frequencies corresponding to $\tau_{off}$ between 4 and 8.5 ms.
Net translation of the particles was observed only within a narrow window of $\tau_{off}=4.7$–6.7ms, in close agreement with the range of 4.6–6.7ms predicted by numerical simulations (Table\ref{tab:results}).
The confinement of directed transport to this $\tau_{off}$ interval is a direct signature of ratcheting, since rectification of Brownian motion can occur only when $\tau_{off}$ lies between the forward and backward diffusion times ($\tau_F$ and $\tau_B$, respectively).
Other possible sources of drift, such as sample tilt, thermal gradients, or concentration inhomogeneities, would instead have yielded a constant displacement, independent of chopping frequency.

Further confirmation of ratcheting was obtained by investigating its dependence on the intensity of the $\excLambda~\mathrm{nm}$ excitation, with chopping frequency set to achieve the optimum $\tau_{off}$ of 5.5 ms.
Because rectification relies on efficient trapping during the potential-on stage, the effect is expected to appear once $|\Delta U| > k_B T$ and to strengthen as the traps become stiffer. 
Consistent with this expectation, no directed motion was observed below $\sim0.5~\mathrm{kW cm^{-2}}$, where the trapping potential was shallower than the thermal energy of the system (Figure~\ref{fig:ratchetting-results}(b)).
Above this threshold, stable trapping was established and ratcheting efficiency increased with intensity up to $\sim0.8~\mathrm{kW cm^{-2}}$, reflecting more effective confinement during the on-phase of the modulation (Figure~\ref{fig:ratchetting-results}(c)).
At higher powers, however, the efficiency decreased and was lost entirely above $1.5~\mathrm{kW cm^{-2}}$.
Temperature measurements under continuous 980 nm illumination showed minimum heating of the dispersion ($<1^\circ$C after 1 hour of heating at $2.4~\mathrm{kW,cm^{-2}}$, see SI), excluding thermophoresis or the increased thermal energy of the system as the cause of the decline.
In fact, trap stiffness increased approximately linearly with excitation power up to $\sim$1.5kWcm$^{-2}$ but above that the rate of increase became markedly steeper, albeit with larger uncertainty (Figure~\ref{fig:ratchetting-results}(c)).
Such behavior is consistent with an increasing fraction of PS particles being driven into contact with the antenna surfaces by stronger near-field gradients, promoting adhesion and thereby reducing overall transport efficiency

To assess the applicability of the plasmonic ratchet beyond the \smallPSsize~nm PS spheres used in its design, we further extended ratcheting experiments to larger PS nanospheres (\(\largePSsize~\mathrm{nm}\)) and to larger nanoparticles of different composition (\(\largePSsize~\mathrm{nm}\), PTB7).
In all cases ratcheting was observed, although at reduced velocities and longer $\tau_{off}$ values compared to the design case (Figure~\ref{fig:ratchetting-results}(d) and (e), Table\ref{tab:results}).
These observations are consistent with the reduced diffusivity and less efficient trapping of the larger nanoparticles: as their size increases, near-field gradients are effectively averaged over the particle volume, yielding stronger forces but shallower and less asymmetric potentials (Figure S6).
Numerical simulations incorporating these effects reproduced the experimentally-observed $\tau_{off}$ ranges over which ratcheting occurred (Table~\ref{tab:results}, Figures S7 and S9).

\begin{table}
 \caption{Effect of nanoparticle size and composition on their ratcheting characteristics.}
 \label{tab:results}
  \begin{tabular}[htbp]{@{}llll@{}}
    \hline
    Particle composition & Polystyrene & Polystyrene & PTB7 \\
    Particle size & \smallPSsize~nm & \largePSsize~nm & \PTBsize~nm \\
    \hline
    Maximum mean velocity, x ($\mu m~s^{-1}$) & 0.14 & 0.06 & 0.15 \\
    Maximum mean velocity, y ($\mu m~s^{-1}$) & 2.37 & 1.61 & 1.84  \\
    Optimum experimental \(t_{off}\), (ms) & 5.5 & 27.7 & 27.7\\
    Theoretical \(t_{off}\) range, (ms): & $4.6 - 6.7$ & $25.5 - 30.8$  & $25.4 - 30.9$\\
    \hline
  \end{tabular}
\end{table}

Minimal differences were observed between PS and PTB7 nanoparticles of comparable size, despite the former being an insulator and the latter a semiconductor with higher polarizability.
Both exhibited ratcheting within nearly identical $\tau_{off}$ ranges (Figure~\ref{fig:ratchetting-results}(d-e)), with PTB7 nanoparticles reaching slightly higher maximum velocities (1.84~$\mu$ms$^{-1}$) than PS spheres (1.55~$\mu$ms$^{-1}$) -- a difference attributable to their marginally smaller size and faster diffusion.
This behavior highlights that net transport is governed primarily by particle diffusivity during the potential-off phase, while variations in polarizability or conductivity -- which alter trap strength -- have little effect.
These results show that plasmonic ratchets function reliably across particle sizes and compositions, with performance ultimately set by the degree of potential asymmetry experienced by the particle.
More broadly, they establish plasmonic ratcheting as a versatile platform for nanoparticle manipulation with potential for wide-ranging applications.

In conclusion, this study demonstrates that plasmonic ratchet systems can provide a highly effective means for the controlled long-range transport of nanoscale particles.
Using arrays of asymmetric plasmonic structures and external optical modulation, a plasmonic ratchet was experimentally-realized and used to achieve efficient rectification of Brownian motion of \smallPSsize ~ - \largePSsize ~nm PS and PTB7 spheres, achieving net transport velocities up to $\maxVel~\mu$m s$^{-1}$ for incident intensities below $\lowPower~\mathrm{kW~cm^{-2}}$.
These results establish the robustness of plasmonic ratcheting across particle sizes and compositions, with performance ultimately governed by particle diffusivity and the asymmetry of the trapping potential.

Compared with other optical ratchet systems based on photonic crystals, holographic tweezers, or scanning laser traps,\cite{wu2016near,lee2005one,faucheux1995optical} which required intensities on the order of $10^{8}$–$10^{14}$ $\mathrm{W}~\mathrm{m^{-2}}$ ($10$-$10^6 ~\mathrm{kW~cm^{-2}}$) to drive nanoscale or microscale particles at modest velocities, the plasmonic approach presented here achieves markedly higher transport speeds at orders-of-magnitude lower power.
Operating at such low intensities mitigates the risks of thermal and photodamage to delicate analytes such as viruses,\cite{burell2017fenner,rath2020investigation} and emphasizes the potential of plasmonic ratchets as a robust, scalable, and energy-efficient solution for the controlled manipulation of nanoscale analytes in lab-on-chip, diagnostic, and nanophotonic applications.

\begin{acknowledgement}

M.P.C., A.M.D. and A.R. acknowledge the funding support from the Royal Society (RGF\textbackslash \\ R1\textbackslash 180068, UF150542, URF\textbackslash R\textbackslash 211023) and from EPSRC (EP/W017075/1). S.A.M. acknowledges the Lee-Lucas Chair in Physics. P.A.H. acknowledges the MICIU and AEI (RYC2021-031568-I and  “María de Maeztu” Programme for Units of Excellence in R\&D, CEX2023-001316-M). D.R. acknowledges ERC iCOMM 789340 for the funding support.

\end{acknowledgement}

\begin{suppinfo}

Supplementary information is available free of charge at NanoLetters journal website.

Principles of Brownian ratchets, details of numerical simulations of optical forces and trapping potentials, fabrication protocols for plasmonic ratchets and PTB7 nanoparticles, particle tracking and data analysis procedures, and additional experimental ratcheting data for PS spheres and PTB7 nanoparticles (PDF).

\end{suppinfo}

\bibliography{achemso-demo}

\end{document}